\documentclass[twocolumn,showpacs,aps,prl,superscriptaddress]{revtex4}

\usepackage{graphicx}
\usepackage{dcolumn}
\usepackage{epsfig}
\usepackage{psfrag}
\usepackage{amsmath}

\newcommand{\BaBarYear}    {08}
\newcommand{\BaBarNumber}  {011}

\newcommand{\SLACPubNumber} {13197}
\newcommand{\LANLNumber} {0804.2422 [hep-ex]}

 \newcommand{\BaBarType}      {PUB}  

\input pubboard/babarsym   

\RequirePackage{xspace}

\newcommand{\pvec}{{\bf p}}

\newcommand{\acp}{\ensuremath{\calA_{ch}}}

\newcommand{\calB}{\ensuremath{{\cal B}}}

\newcommand{\timemsix}{\ensuremath{\times10^{-6}}}

\newcommand{\DE}{\ensuremath{\Delta E}}

\newcommand{\xf}{\ensuremath{{\cal F}}}
\newcommand{\hel}{\ensuremath{{\cal H}}}
\newcommand{\thetaT}{\ensuremath{\theta_{\rm T}}}
\newcommand{\costhr}{\ensuremath{\cos\thetaT}}

\newcommand\etal{{\it et al.}}
\newcommand{\half}{\ensuremath{\frac{1}{2}}}

\newcommand{\bma}[1]{\boldmath{$#1$}}

\newcommand{\bfig}{\begin{figure}[htbpc!]}
\newcommand{\efig}{\end{figure}}
\newcommand\bef{\begin{figure}}
\newcommand\edf{\end{figure}}
\newcommand\dbline{\noalign{\vskip 0.10truecm\hrule}\noalign{\vskip 2pt}\noalign{\hrule\vskip 0.10truecm}}
\providecommand{\tbline}{\noalign{\vskip 0.05truecm\hrule\vskip0.05truecm}}
\newcommand\sgline{\noalign{\vskip 0.10truecm\hrule\vskip 0.10truecm}}
\newcommand\beq{\begin{equation}}
\newcommand\eeq{\end{equation}}
\newcommand\bear{\begin{array}}
\newcommand\enar{\end{array}}
\newcommand\beqa{\begin{eqnarray}}
\newcommand\eeqa{\end{eqnarray}}
\newcommand\ben{\begin{enumerate}}
\newcommand\een{\end{enumerate}}

\newcommand{\UfourS}{\ensuremath{\Upsilon(4S)}}

\newcommand{\etagg}{\ensuremath{\eta_{\gaga}}}
\newcommand{\etappp}{\ensuremath{\eta_{3\pi}}}

\newcommand{\etapepp}{\ensuremath{\etapr_{\eta\pi\pi}}}

\newcommand{\etaprg}{\ensuremath{\etapr_{\rho\gamma}}}

\newcommand{\Kst}{\ensuremath{K^*}}

   \newcommand{\rhop}{\ensuremath{\rho^+}}
   
   \newcommand{\rhoz}{\ensuremath{\rho^0}}

\newcommand{\fetapiz}{\ensuremath{\eta\piz}\xspace}
\newcommand{\etapiz}{\ensuremath{\Bz\ra\fetapiz}\xspace}
\newcommand{\Betapiz}{\ensuremath{\calB(\etapiz)}\xspace}
\newcommand{\retapiz}{\ensuremath{xx^{+xx}_{-xx}\pm xx}\xspace}

\newcommand{\uletapiz}{\ensuremath{xx}\xspace}

\newcommand{\setapiz}{\ensuremath{xx}\xspace}
   \newcommand{\fetaggpiz}{\ensuremath{\eta_{\gaga} \piz}\xspace}
   
   \newcommand{\fetappppiz}{\ensuremath{\eta_{3\pi} \piz}\xspace}

\newcommand{\fetarhop}{\ensuremath{\eta\rho^+}}
\newcommand{\etarhop}{\ensuremath{\Bp\ra\fetarhop}}
\newcommand{\Betarhop}{\ensuremath{\calB(\etarhop)}}
\newcommand{\retarhop}{\ensuremath{xx^{+xx}_{-xx}\pm xx}}
\newcommand{\Retarhop}{\ensuremath{(\retarhop)\times 10^{-6}}}

\newcommand{\Aetarhop}{\ensuremath{xx\pm xx \pm xx}}
\newcommand{\setarhop}{\ensuremath{xx}}
  \newcommand{\fetaggrhop}{\ensuremath{\eta_{\gamma\gamma} \rho^+}}
  \newcommand{\etaggrhop}{\ensuremath{\Bp\ra\fetaggrhop}}
  \newcommand{\fetappprhop}{\ensuremath{\eta_{3\pi} \rho^+}}

\newcommand{\fetappiz}{\ensuremath{\etapr\piz}\xspace}
\newcommand{\etappiz}{\ensuremath{\Bz\ra\fetappiz}\xspace}
\newcommand{\Betappiz}{\ensuremath{\calB(\Bz\ra\etapr \piz)}\xspace}
\newcommand{\retappiz}{\ensuremath{xx^{+xx}_{-xx} \pm xx}\xspace}

\newcommand{\uletappiz}{\ensuremath{xx}\xspace}

\newcommand{\setappiz}{\ensuremath{xx}\xspace}
   \newcommand{\fetapepppiz}{\ensuremath{\etapr_{\eta\pi\pi} \piz}\xspace}
   
   \newcommand{\fetaprgpiz}{\ensuremath{\etapr_{\rho\gamma} \piz}\xspace}

\newcommand{\fetapeta}{\ensuremath{\etapr\eta}}
\newcommand{\etapeta}{\ensuremath{\Bz\ra\fetapeta}}
\newcommand{\Betapeta}{\ensuremath{\calB(\etapeta)}}
\newcommand{\retapeta}{\ensuremath{xx^{+xx}_{-xx}\pm xx}}

\newcommand{\uletapeta}{\ensuremath{xx}}

\newcommand{\setapeta}{\ensuremath{xx}}
  \newcommand{\fetaprgetagg}{\ensuremath{\etaprg\etagg}}
  \newcommand{\fetapeppetagg}{\ensuremath{\etapepp\etagg}}
  \newcommand{\fetaprgetappp}{\ensuremath{\etaprg\etappp}}
  \newcommand{\fetapeppetappp}{\ensuremath{\etapepp\etappp}}

  \newcommand{\etaprgetagg}{\ensuremath{\Bz\ra\fetaprgetagg}}

\newcommand{\fomegapiz}{\ensuremath{\omega\pi^0}\xspace}
\newcommand{\omegapiz}{\ensuremath{\Bz\ra\fomegapiz}\xspace}
\newcommand{\Bomegapiz}{\ensuremath{\calB(\omegapiz)}\xspace}
\newcommand{\romegapiz}{\ensuremath{xx\pm xx \pm xx}\xspace}

\newcommand{\ulomegapiz}{\ensuremath{xx}\xspace}

\newcommand{\somegapiz}{\ensuremath{xx}\xspace}

\renewcommand{\retarhop}{\ensuremath{9.9\pm 1.2\pm 0.8}}
\renewcommand{\Aetarhop}{\ensuremath{0.13\pm 0.11 \pm 0.02}}
\renewcommand{\setarhop}{\ensuremath{9.0}}

\renewcommand{\retapiz}{\ensuremath{0.9\pm0.4\pm 0.1}}		
\renewcommand{\uletapiz}{\ensuremath{1.5}}			
\renewcommand{\setapiz}{\ensuremath{2.2}}			

\renewcommand{\retappiz}{\ensuremath{0.9\pm0.4\pm 0.1}}		
\renewcommand{\uletappiz}{\ensuremath{1.5}}			
\renewcommand{\setappiz}{\ensuremath{3.1}}			

\renewcommand{\romegapiz}{\ensuremath{0.07\pm 0.26\pm 0.02}} 
\renewcommand{\ulomegapiz}{\ensuremath{0.5}}			
\renewcommand{\somegapiz}{\ensuremath{0.3}}			

\renewcommand{\retapeta}{\ensuremath{0.5\pm0.4\pm 0.1}}		
\renewcommand{\uletapeta}{\ensuremath{1.2}}			
\renewcommand{\setapeta}{\ensuremath{1.4}}			

\newcommand{\theTitle}{{\large \bf\boldmath
Observation of \etarhop\ and search for \Bz\ Decays
to \fetapeta, \fetapiz, \fetappiz, and \fomegapiz
}}

\begin{document}



 \begin{flushleft}
 \babar-\BaBarType-\BaBarYear/\BaBarNumber \\
 SLAC-PUB-\SLACPubNumber \\
 arXiv:\LANLNumber
 \end{flushleft}

\title{
 \theTitle
}

%
\author{B.~Aubert}
\author{M.~Bona}
\author{Y.~Karyotakis}
\author{J.~P.~Lees}
\author{V.~Poireau}
\author{E.~Prencipe}
\author{X.~Prudent}
\author{V.~Tisserand}
\affiliation{Laboratoire de Physique des Particules, IN2P3/CNRS et Universit\'e de Savoie, F-74941 Annecy-Le-Vieux, France }
\author{J.~Garra~Tico}
\author{E.~Grauges}
\affiliation{Universitat de Barcelona, Facultat de Fisica, Departament ECM, E-08028 Barcelona, Spain }
% ...
\collaboration{The \babar\ Collaboration}
\noaffiliation

\date{\today}

\begin{abstract}
We present measurements of branching fractions for five $B$-meson decays to 
two-body charmless final states.  The data, collected with the
\babar\ detector at the Stanford Linear Accelerator Center, represent
459 million \BB\ pairs. The results for branching fractions are, in
units of $10^{-6}$ (upper limits at 90\% C.L.): 
$\Betarhop = \retarhop$,
$\Betapeta = \retapeta\ (<\uletapeta)$,
$\Betapiz = \retapiz\ (<\uletapiz)$, 
$\Betappiz = \retappiz\ (<\uletappiz)$, and
$\Bomegapiz = \romegapiz\ (<\ulomegapiz)$.
The first error quoted is statistical and the second systematic.  For the 
\fetarhop\ mode, we measure the charge asymmetry $\acp(\etarhop)=\Aetarhop$.
\end{abstract}


\maketitle

Measurements of charmless $B$ decays are now routinely used to test the 
accuracy of theoretical predictions based on, for example, QCD
factorization~\cite{SUthreeQCDFact,acpQCDfact}, flavor~SU(3)
symmetry~\cite{FUglob,chiangPP,chiangGlob}, perturbative QCD~\cite{pQCD}, or
soft collinear effective theory \cite{SCET}.  We present measurements of
the branching fraction and charge asymmetry for the decay \etarhop\
(charge conjugate reactions are implied throughout this paper), superseding 
our previous result that found a 4.7$\sigma$ signal for this decay \cite{BABARetarhop} with a
luminosity of about one-half that used in this paper.  In addition we
search for the decays \etapeta, \etapiz, \etappiz, and \omegapiz.
None of these decays have been observed previously though limits 
have been reported by \babar\ \cite{BABARPRD}, Belle \cite{BellePUB},
and CLEO \cite{CLEOPUB}.

In the Standard Model (SM) the dominant processes that contribute to these 
decays are described by tree amplitudes and to a lesser extent 
penguin (loop) amplitudes.  For \etappiz\ and \etapiz\ the color-suppressed 
tree diagram is suppressed by approximate cancellation between the amplitudes 
for the \piz\ and for the isoscalar meson that contains the spectator quark.  
The approximate ranges of expectations
\cite{SUthreeQCDFact,acpQCDfact,FUglob,chiangPP,chiangGlob,pQCD,SCET}
for the branching fraction are $\sim$10$\times10^{-6}$ for
\etarhop, $0.3$--2\timemsix\ for \etapeta, 0.2--1.0\timemsix\ for
$\Bz\ra\eta^{(\prime)}\piz$, and $\sim$0.1\timemsix\ for $\Bz\ra\omega\piz$.
Direct \CP\ violation could be detected as a charge asymmetry, defined as
$\acp \equiv (\Gamma^--\Gamma^+)/(\Gamma^-+\Gamma^+)$, where the superscript
on the width $\Gamma$ corresponds to the sign of the $B^\pm$ meson; \acp\ for 
\etarhop\ is expected to be small since the decay is dominated by a single 
amplitude.

These \Bz\ decays are also of interest in constraining the expected value of
the time-dependent \CP-violation asymmetry parameter $S_f$ in the $B$ decay 
with final state $f=\etapr\KS$~\cite{chiangGlob,GLNQ03,GRZ06}.  
The leading-order SM calculation gives the equality
$S_{\etapr\KS} = S_{J/\psi\KS}$, where the latter has been
precisely measured \cite{sin2beta}, and 
equals \stwob\ in the SM. The \CP\ asymmetries in the charmless $B$
decays are not only sensitive to contributions from new physics, but also to
contamination from sub-leading SM amplitudes. Recent theoretical calculations
of the size of the change in $S_{\etapr\KS}$ from these sub-leading amplitudes 
finds no more than 0.03 \cite{Beneke,SCET}.  The most stringent constraint 
from data on such contamination uses SU(3) and the measured
branching fractions of the decays \etapeta, \etapiz, \etappiz\
\cite{chiangGlob,GLNQ03,GRZ06}.  Recently it has also
been suggested~\cite{GZ05,G05} that
\etappiz\ and \etapiz\ can be used to constrain the
contribution from isospin-breaking effects on the value of \stwoa\ in
$B\to\pi^+\pi^-$ decays.

The results presented here are based on data collected with the \babar\
detector~\cite{BABARNIM} at the PEP-II asymmetric $e^+e^-$
collider located at the Stanford Linear Accelerator Center.
We recorded a data sample at the $\Upsilon (4S)$ resonance
(center-of-mass energy $\sqrt{s}=10.58\ \gev$) with an integrated luminosity 
of 418 \invfb, corresponding to $(459\pm5)\times 10^6$ \BB\ pairs.

Charged particles from the \epem\ interactions are detected and their
momenta measured by a combination of five layers of double-sided
silicon microstrip detectors and a 40-layer drift chamber,
both operating in the 1.5~T magnetic field of a superconducting
solenoid. Photons and electrons are identified with a CsI(Tl)
electromagnetic calorimeter (EMC).  Further charged particle
identification (PID) is provided by the average energy loss ($dE/dx$) in
the tracking devices and by an internally reflecting ring imaging
Cherenkov detector (DIRC) covering the central region.

We establish the event selection criteria with the aid of a detailed
Monte Carlo (MC) simulation of the \B\ production and decay sequences,
and of the detector response \cite{geant}.  These criteria are designed
to retain signal events with high efficiency.  Applied to the data, they
result in a sample
much larger than the expected signal, but with well characterized
backgrounds. We extract the signal yields from this sample with a
maximum likelihood (ML) fit.

The \B-daughter candidates are reconstructed through their decays
$\piz\ra\gaga$, $\eta\ra\gaga$ (\etagg), $\eta\ra\pip\pim\piz$ (\etappp), 
$\etapr\ra\etagg\pip\pim$ (\etapepp), $\etapr\ra\rhoz\gamma$ (\etaprg), 
$\omega\ra\pip\pim\piz$, $\rhoz\ra\pip\pim$ and $\rhop\ra\pip\piz$.
Table \ref{tab:rescuts}\ lists the
requirements on the invariant masses of these particles' final states.
All requirements are kept loose ($>3\sigma$) for later fitting except
for the \piz\ invariant mass which is not included in the fits.
Secondary charged pions in $\eta$, \etapr, and $\omega$ candidates are rejected
if classified as protons, kaons, or electrons by a combination of their DIRC, 
$dE/dx$, and EMC PID signatures. 

\begin{table}[!hbtp]
\begin{center}
\caption{
Selection requirements on the invariant masses of resonances and the
laboratory energies of photons from their decay.}
\label{tab:rescuts}
\begin{tabular}{lcc}
\dbline
State		& Invariant mass (MeV)		& $E(\gamma)$ (MeV)\\
\sgline						
Prompt \piz		& $120 < m(\gamma\gamma) < 150$		& $>50$		\\
\etagg		& $490 < m(\gamma\gamma) < 600$		& $>100$	\\
\etappp		& $520 < m(\pip\pim\piz) < 570$		& $>30$		\\
\etapepp	& $910 < m(\pip\pim\eta) <1000$		& $>100$	\\
\etaprg		& $910 < m(\pip\pim\gamma) <1000$	& $>200$	\\
$\omega$	& $735 < m(\pip\pim\piz) < 825$		& $>30$		\\
\rhoz		& $510 < m(\pip\pim) <1000$		& ---	\\
\rhop		& $470 < m(\pip\piz) <1070$		& $>30$		\\
\dbline
\end{tabular}
\vspace{-5mm}
\end{center}
\end{table}

We reconstruct the \B-meson candidate by combining the
four-momenta of a pair of daughter mesons, with a vertex constraint if
the ultimate final state includes at least two charged particles.  Since
the natural widths of the $\eta$, \etapr, and \piz\ mesons are much smaller
than the resolution, we also constrain their masses to nominal values
\cite{PDG2006}\ in the fit of the \B\ candidate.
From the kinematics of the \UfourS\ decay we determine the energy-substituted mass
$\mes=\sqrt{\frac{1}{4}s-\pvec_B^2}$
and the energy difference $\DE = E_B-\half\sqrt{s}$, where
$(E_B,\pvec_B)$ is the $B$-meson 4-momentum vector, and
all values are expressed in the \UfourS\ rest frame.
The resolution in \mes\ is 3.0 MeV and in \DE\ is 24--50 MeV, depending
on the decay mode.  We
require 5.25\ \gev$<\mes<5.29$ GeV and $|\DE|<0.25$ GeV 
($<0.2$ GeV for \etapeta\ and \etarhop).

Backgrounds arise primarily from random combinations of particles in
continuum $\epem\ra\qqbar$ events ($q=u,d,s,c$).  We reduce these with
requirements on the angle \thetaT\ between the thrust axis of the $B$ 
candidate's decay products in the \UfourS\ rest frame and the thrust axis of the rest of 
the charged tracks and neutral calorimeter clusters in the event.  The 
distribution is sharply peaked near $|\costhr|=1$ for \qqbar\ jet pairs
and nearly uniform for $B$-meson decays.  We require
$|\costhr|<0.7$--$0.9$ depending on the decay mode.  

In the ML fit we discriminate against \qqbar\ background with a
Fisher discriminant \xf\ that combines five variables \cite{omegaPRD}: 
the polar angles, with respect to the beam axis in the \UfourS\ rest 
frame, of the $B$ candidate momentum and of the $B$ thrust axis; 
the flavor tagging category \cite{ccbarK0}; and the zeroth and second 
angular moments of the energy flow, excluding the $B$ candidate, about the 
$B$ thrust axis.
It provides about one standard deviation of separation between \B\ decay 
events and combinatorial background.

We also impose restrictions on decay angles to exclude the most asymmetric
decays where soft-particle backgrounds concentrate and the acceptance
changes rapidly.  We define the decay angle $\theta_{\rm dec}^k$ and its
cosine $\hel_k$ for a meson
$k$ as the angle between the momenta of a daughter particle and the meson's
parent, measured in the meson's rest frame.  We require for the \etaprg\
decays $|\hel_{\rhoz}|< 0.9$ and for $\eta^{(\prime)}\piz$
$|\hel_{\piz}| < 0.95 $.  For \etaprgetagg\ we suppress the 
background $B\ra\Kst\gamma$ by requiring $|\hel_{\eta}|< 0.86$.  These
distributions are uniform for signal except for $\hel_{\rhoz}$ which has 
a $1-\hel_\rho^2$ distribution.

For the \etarhop\ decay, we define $\theta_{\rm dec}^k$ as the angle 
between the \piz\ and the negative of the $B$ momentum in the $\rho$ rest frame.
We require $-0.75<\hel_{\rhop}<0.95$.
For the \omegapiz\ decay, $|\hel_\omega|$ is defined as the cosine of the
angle between the normal to the $\omega$ decay plane (the plane of 
the three pions in the $\omega$ rest frame) and the
flight direction of the $B$, measured in the $\omega$ rest frame.
Both of these quantities have a $\hel^2$ distribution for signal.

The average number of candidates found per selected event is in the range
1.06 to 1.47, depending on the final state.  We choose the candidate with
the largest probability for the fit to the $B$ decay tree.

We obtain yields for each channel from a ML fit with the input observables 
\DE, \mes, \xf, $m_k$, $k=1,2$ (the daughter invariant mass spectrum of the 
$\eta$, \etapr, $\omega$, or \rhop\ candidate), and $\hel_k$ the helicity of the
$\omega$ or \rhop\ candidate.  The selected sample sizes are given in the second
column of Table~\ref{tab:results}.  Besides any signal events, the samples
contain combinatorial background from \qqbar\ (dominant) and \BB\ with $b\ra c$,
and a component from other charmless \BB\ modes that we estimate from the 
simulation to be no more than two percent of the sample.  The latter events have
ultimate final states different from the signal, but with similar kinematics 
so that broad peaks near those of the signal appear in some observables,
requiring a separate component in the probability density function (PDF).

The likelihood function is
\begin{eqnarray}
{\cal L} = \exp{(-\sum_{j} Y_{j})} \prod_i^{N}\sum_{j} 
Y_{j} \times~~~~~~~~~~~~~~~~~~~~~~~~~~~~~~~\label{eq:likelihood}\\
{\cal P}_j(\mes^i){\cal P}_j(\DE^i){\cal P}_j(\xf^i){\cal P}_j (m_1^i)
\left[{\cal P}_j(m_2^i),{\cal P}_j(\hel_{\omega,\rhop}^i)\right], \nonumber
\end{eqnarray}
where $N$ is the number of events in the sample, and, for each of the three 
components $j$, $Y_{j}$ is the yield of events and ${\cal P}_j(x^i)$ the
PDF for observable $x$ in event $i$. For the mode 
$\Bz\ra\etapepp\etappp$ we found no need for the charmless \BB\ background
component.   For the \etaggrhop\ and \omegapiz\ decays we split the charmless 
\BB\ PDF into components made from backgrounds with and without a \rhop.
The factored form of the PDF indicated in Eq.\
\ref{eq:likelihood}\ is appropriate since correlations among observables
measured in the data are small.  Distortions of the signal yields caused 
by this approximation are measured in simulation and
included in the bias corrections and systematic errors discussed below.

We determine the PDFs for the signal and charmless \BB\ background components
from fits to MC simulated events.  Large control samples of $B$ decays to 
charmed final states of similar topology [$\Bp\ra\Dzb(\Kp\pim\piz)\pip$ 
and $\Bp\ra\Dzb(\Kp\pim\piz)\rhop$] are used to verify the simulated 
resolutions in \DE\ and \mes.  Where the control data samples reveal small 
differences from MC, we shift or scale the resolution used in the ML fits.
We develop PDFs for
the combinatorial background with fits to the data from which the signal
region ($5.27\ \gev<\mes<5.29\ \gev$ and $|\DE|<0.1$ GeV) has been excluded.

\begin{table*}[!hbtp]
\caption{
Number of events $N$ in the sample, fitted signal yield $Y_S$ in events (ev.)
with statistical error, measured bias, detection efficiency $\epsilon$, 
daughter branching fraction product ($\prod\calB_i$), and measured branching 
fraction \calB\ with statistical error for each decay chain, and the
measured charge asymmetry \acp\ for the decay \etarhop.
For the combined measurements the significance~$\cal S$
(with systematic uncertainties included), branching fraction with
statistical and systematic error, and in parentheses the 90\%
C.L. upper limits.
}
\label{tab:results}
\newcommand{\mn}{\ensuremath{\phantom{-}}}
\newcommand{\eff}{$\epsilon$ (\%)}
\newcommand{\pbf}{$\prod\calB_i$ (\%)}
\newcommand{\signf}{$\cal S$ ($\sigma$)}
\begin{tabular}{lrclccccc}
\dbline
Mode  & $N$ (ev.) &$Y_S$ (ev.) & Bias (ev.)&\eff &\pbf &\signf&\multicolumn{1}{c}{\calB\ $(10^{-6})$}& \acp\\
\tbline
\bma{\fetarhop}	&  & & &  & &\bma{\setarhop}	&\bma{\retarhop}	&\bma{\Aetarhop}    \\
~~\fetaggrhop	&104609&$326^{+44}_{-42}$&$17\pm9$ &16.7&39.4&    &$10.2\pm1.4$&$0.07\pm0.12$    \\
~~\fetappprhop	&47918&$123^{+27}_{-26}$& $13\pm7$ &11.7&22.6&    &$\mn9.1\pm2.2$&$0.28\pm0.21$    \\
\bma{\fetapeta}		& 	&	& 	&	&	& {\boldmath \setapeta}	&{\boldmath \retapeta} \quad({\boldmath $<$\uletapeta})	\\ 
~~\fetapeppetagg	& 2191	& $8.8^{+6.4}_{-5.1}$	& $0.9\pm0.5$ & 25.6	&  6.9	& 	& $\mn1.0^{+0.8}_{-0.6}$ \\
~~\fetapeppetappp	& 896	& $3.2^{+5.1}_{-4.1}$	& $0.2\pm0.2$ & 16.7	& 4.0	& 	& $\mn1.0^{+1.7}_{-1.4}$ \\
~~\fetaprgetagg		& 39723	& $0.7^{+12.2}_{-8.6}$	& $0.0\pm0.5$ & 25.6	& 11.6	& 	& $\mn0.1^{+0.7}_{-0.6}$ \\
~~\fetaprgetappp	& 20672 & $0.7^{+9.4}_{-6.8}$	& $2.0\pm1.0$ & 18.2	&  6.7	&	& $-0.2^{+1.7}_{-1.2}$ 	\\
\bma{\fetapiz}		& 	&	& 	&	& & {\boldmath \setapiz}	& {\boldmath \retapiz} \quad({\boldmath $<$\uletapiz})	\\
~~\fetaggpiz		& 9085	& $18.6^{+23.9}_{-21.7}$& $4.4\pm2.2$ & 20.5	& 39.4	& 	& $\mn0.4\pm0.6$ 	\\
~~\fetappppiz		& 4030	& $23.3^{+12.5}_{-11.1}$& $1.7\pm0.9$ & 17.3	& 22.6	& 	& $\mn1.3^{+0.7}_{-0.6}$ 	\\
\bma{\fetappiz}		&     	&	&  	&     	&  & {\boldmath \setappiz}	& {\boldmath \retappiz} \quad({\boldmath $<$\uletappiz})	\\
~~\fetapepppiz		& 3784	& $20.6^{+9.4}_{-8.0}$	& $1.8\pm0.9$ & 22.5	& 17.5	&	& $\mn1.1^{+0.5}_{-0.4}$ 	\\
~~\fetaprgpiz		&19789	& $12.2^{+18.4}_{-16.3}$& $2.7\pm1.4$ & 18.9	& 29.5	&	& $\mn0.4^{+0.7}_{-0.6}$  \\
\bma{\fomegapiz}	&39822	& $2.4^{+19.9}_{-16.8}$	& $0.5\pm0.5$ & 18.4	& 89.1	& {\boldmath \somegapiz}	& {\boldmath \romegapiz} \quad({\boldmath $<$\ulomegapiz})	\\
\dbline
\end{tabular}
\end{table*}

\begin{figure*}[!htbp]
\psfrag{FFF}{{\LARGE$\cal F~~~~$}}
\scalebox{.45}[.45]{\includegraphics{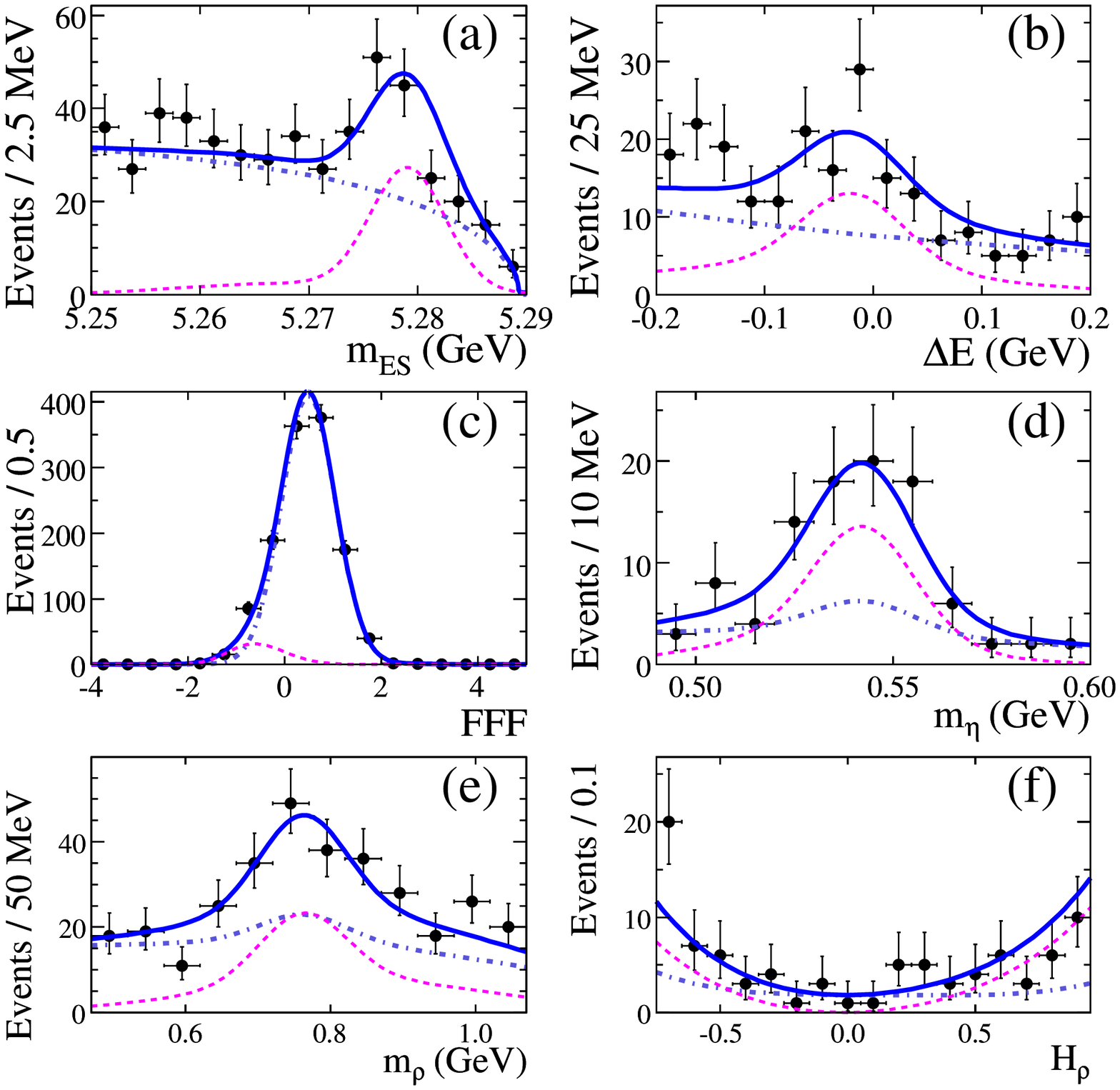}\includegraphics{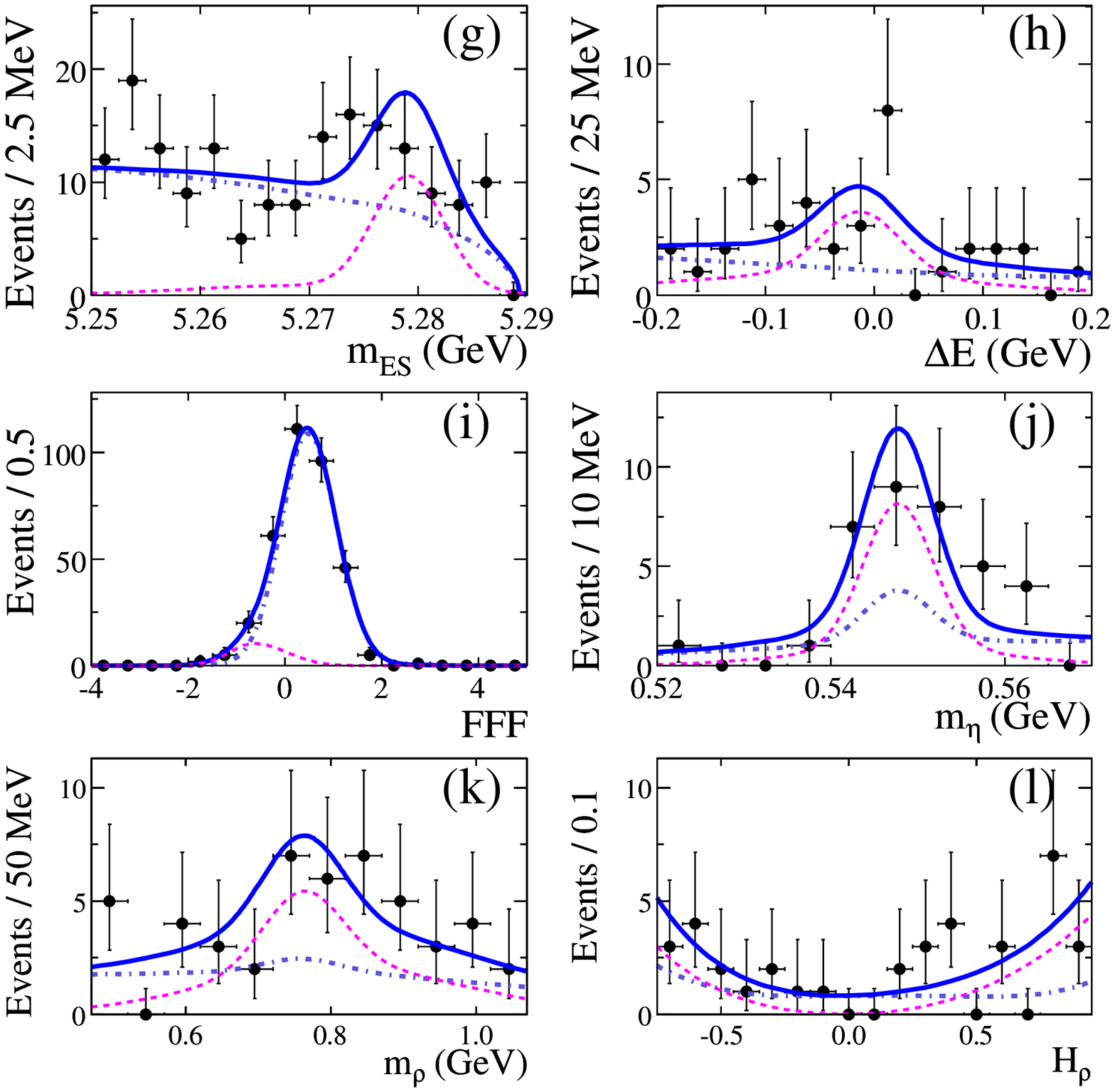}}
\caption{Signal-enhanced projections for \fetaggrhop\ (left) and \fetappprhop\
(right) for (a,g) \mes, (b,h) \DE, (c,i) \xf, (d,j) $m_\eta$, (e,k) $m_\rho$,
(f,l) $\hel_\rho$.  The total (\qqbar\ plus \BB) background fit function is 
shown as blue dot-dashed, signal as magenta dashed, 
and the total as a solid blue line.  These plots are made with a minimum
requirement on the likelihood that has an efficiency for signal of 15--35\%
while reducing the background by between two and three orders of magnitude.}
\label{fig:etarhop_proj}
\end{figure*}

We use the following functional forms for the PDFs: sum of
two Gaussians for ${\cal P}_{\rm sig}(\mes)$, ${\cal P}_{{\rm sig},\BB}(\DE)$,
and the sharper structures in ${\cal P}_{\BB}(\mes)$ and ${\cal
P}_j(m_k)$; linear or quadratic dependences for 
combinatorial components of ${\cal P}_{\BB,\qqbar}(m_k)$ and for ${\cal
P}_{\qqbar}(\DE)$; quadratic functions for ${\cal P}_j(|\hel_{\omega}|)$ and
${\cal P}_j(\hel_{\rhop})$; and a Gaussian of different widths below and 
above the peak, plus a broad Gaussian, for ${ \cal P}_j(\xf)$.  The \qqbar\ 
background in \mes\ is described by the function
$x\sqrt{1-x^2}\exp{\left[-\xi(1-x^2)\right]}$, with
$x\equiv2\mes/\sqrt{s}$ and parameter $\xi$.  These are discussed in
more detail in Ref.~\cite{PRD04} and can be seen in Fig.~\ref{fig:etarhop_proj}
for the \etarhop\ decay.

We allow the parameters most important for the determination of the
combinatorial background PDFs to vary in the fit, along with the yields
for all components.  Specifically, the free background parameters are
most or all of the following, depending on the decay mode: $\xi$ for
\mes, linear and quadratic coefficients for \DE, area and slope of the
combinatorial component for $m_k$, and the mean, width, and width
difference parameters for \xf.  Results for the signal yields are presented in
the third column of Table \ref{tab:results}\ for each sample.

\begin{figure}[!htb]
 \includegraphics[angle=0,width=\linewidth]{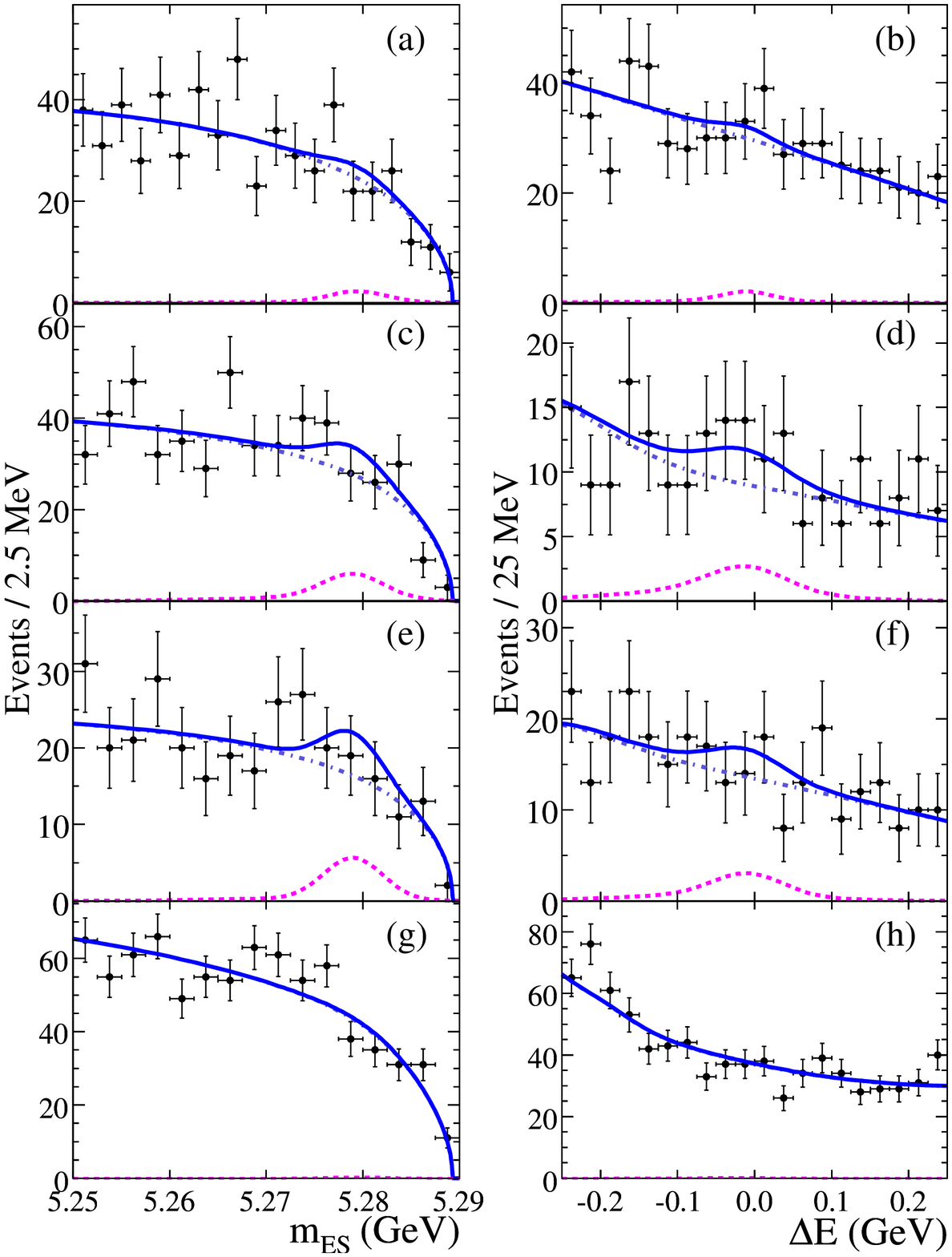}
 \caption{\label{fig:projetapi0}
Signal-enhanced projections of the \Bz-candidate \mes\ and \DE\ mass for (a, b) 
\etapeta, (c, d) \etapiz, (e, f) \etappiz, and (g, h) \omegapiz.  
Points with errors represent data, solid curves the full fit functions (both
signal modes combined), dot-dashed curves the background functions (the 
peaking \BB\ background component is small), and dashed curve the fit
signal function.  These plots are made with a minimum requirement on the 
likelihood that has an efficiency for signal of 45--75\%.
}
\end{figure}

We test and calibrate the fitting procedure by applying it to ensembles of 
simulated experiments composed of \qqbar\ events drawn from the PDF, into which
we have embedded the expected number of signal and charmless \BB\ background
events randomly extracted from the fully simulated MC samples. We find
biases of 0--17 events, somewhat dependent on the signal yield.  The bias
values obtained for simulations that reproduce the yields found in the
data are given in the fourth column of Table~\ref{tab:results}.
Figure \ref{fig:etarhop_proj} shows PDFs and projections of subsamples
of the data, enriched with a threshold requirement on the signal likelihood 
(computed without the variable plotted) for the \etarhop\ decay.  
Figure \ref{fig:projetapi0} shows projections for the other four modes.

We determine the reconstruction efficiencies, given in Table
\ref{tab:results}, as the ratio of reconstructed and accepted events in
simulation to the number generated.  We compute the branching fraction for 
each channel by subtracting the fit bias from the measured yield, and 
dividing the result by the efficiency and the number of produced \BB\ 
pairs~\cite{PRD04}.  We assume that the branching fractions of the \UfourS\ 
to \BpBm\ and \BzBzb\ are each equal to 50\%.
Table \ref{tab:results} gives the numbers pertinent to these computations.

We combine results where we have multiple decay channels by adding 
for each channel the function
$-2\ln{\left\{\left[\calL(\calB)/\calL(\calB_0)\right]\otimes
G(\calB;0,\sigma^\prime)\right\}}$, where $\calB_0$ is the central value 
from the fit, $\sigma^\prime$ is the part of the systematic uncertainty
uncorrelated with other channels, and $\otimes G$ denotes convolution with a 
Gaussian function; the part of the systematic
uncertainty common to all channels is then added in quadrature.
We give the resulting final branching fractions for each mode in 
Table \ref{tab:results} with the significance, taken as the square root 
of the difference between the value of $-2\ln{\cal L}(\calB)$ (with only 
additive systematic uncertainties included) for zero signal and the value at 
its minimum.  The 90\%\ confidence level (C.L.) upper limits are taken to be 
the branching fraction below which lies 90\% of the total of the likelihood
integral in the positive branching fraction region.

The systematic uncertainties on the branching fractions arising from 
lack of knowledge of the PDFs have been included in part in the statistical 
error since most background parameters are free in the fit.  For the
signal, the uncertainties in PDF parameters are estimated from the
consistency of fits to MC and data in control modes.  Varying the
signal-PDF parameters within these errors, we estimate yield
uncertainties of 0.3--7 events, depending on the decay mode.  The uncertainty
from fit bias (Table \ref{tab:results}) includes the statistical uncertainty 
from the simulated experiments added in quadrature with one-half of the 
fit-bias correction.  We estimate the
uncertainty from modeling the charmless \BB\ backgrounds by accounting for the
uncertainties in the knowledge of their branching fractions.
These additive errors are the largest systematic errors for the modes
with small signal yield (but not \etarhop).

Uncertainties in our knowledge of the efficiency, found from auxiliary
studies, include $0.4\%\times N_t$ and $1.5\%\times N_\gamma$, where
$N_t$ and $N_\gamma$ are the number of tracks and photons, respectively,
in the $B$ candidate.  The uncertainty in the total number of \BB\ pairs in the
data sample is 1.1\%.  Published data \cite{PDG2006}\ provide the
uncertainties in the $B$-daughter product branching fractions (0.7--3.9\%).
The uncertainties in the efficiency from the event selection are about 0.5\%.

We observe the decay \etarhop\ with a significance of 9 standard
deviations.  The branching fraction \Retarhop\ and charge asymmetry
\Aetarhop\ are in good agreement with
the theoretical expectations.  We do not find evidence for the other four 
decays though the sensitivity of these measurements is now comparable to the
range of the theoretical estimates.  

We are grateful for the excellent luminosity and machine conditions
provided by our \pep2\ colleagues, 
and for the substantial dedicated effort from
the computing organizations that support \babar.
The collaborating institutions wish to thank 
SLAC for its support and kind hospitality. 
This work is supported by
DOE
and NSF (USA),
NSERC (Canada),
CEA and
CNRS-IN2P3
(France),
BMBF and DFG
(Germany),
INFN (Italy),
FOM (The Netherlands),
NFR (Norway),
MES (Russia),
MEC (Spain), and
STFC (United Kingdom). 
Individuals have received support from the
Marie Curie EIF (European Union) and
the A.~P.~Sloan Foundation.

%

\renewcommand{\baselinestretch}{1}

\end{document}